# TOWARDS EFFICIENT CALCULATION OF INFORMATION MEASURES FOR REORDERING OF BINARY DECISION DIAGRAMS

*Denis V. Popel*[*]

Department of Computer Science, University of Wollongong,
Dubai Campus, P.O. Box 20183, Dubai, U.A.E.
popel@ieee.org

## ABSTRACT

This paper introduces new technique for efficient calculation of different Shannon information measures which operates *Binary Decision Diagrams* (BDDs). We offer an algorithm of BDD reordering which demonstrates the improvement of the obtaining outcomes over the existing reordering approaches. The technique and the reordering algorithm have been implemented, and the results on circuits' benchmarks are analyzed. We point out that the results are quite promising, the algorithm is very fast, and it is easy to implement. Finally, we show that our approach to BDD reordering can yield to reduction in the power dissipation for the circuits derived from BDDs.

## 1. INTRODUCTION

Over the years, many important problems in digital circuits synthesis and optimization have been approached using concepts from *Information Theory*. In the different fields of Computer Aided Design (CAD) of digital circuits, one is often faced with the task of computing information measures, especially Shannon entropy, of each of the signal leads in the network. It is quite natural and valuable to consider switching activity in the terms of entropy measures and to use entropy as a complexity characteristic of optimization process [7]. For instance, the dominant part of power dissipation in integrated circuits is the dynamic power dissipation which is directly proportional to the signal switching activity. The problem of estimating switching activity could be solved efficiently if fast procedures for probabilities calculation are available [1]. In our research, we use the approach to compute probabilities based on the idea that the digital circuit is considered as a network with the probabilities of signal occurrence assigned to internal connections and outputs [3, 6].

The objective of this paper is to develop novel BDD based technique for calculation of different information measures for further application in CAD problems. Additionally, we present an algorithm of BDD reordering which can be applied for the synthesis of low-power circuits derived from BDDs. Among many emerging approaches, digital circuits derived from BDDs are of great interest to engineers and scientists as they compound the junction for the class of low-power dissipation and highly testable circuits. Concepts and techniques developed for circuits derived from BDDs can therefore be extended to other classes of digital circuits. On the whole, results obtained in the study of appliance of information theory may offer useful clues as how to tackle optimization problems concerning the design of digital circuits.

The paper is structured as follows. In Section 2, we briefly review the basic notations of BDDs and information theory. Section 3 introduces a new technique to calculate information measures taking into consideration the assignment of probabilities on the graph representation. Then, we point out applications of this technique to BDD reordering and demonstrate case-study results in Section 4. Section 5 describes the conclusions.

## 2. PRELIMINARIES AND ASSUMPTIONS

Consider logic representation of a digital circuit in the form of *Boolean function* $f$ treated as the mapping $\mathbf{B}^n \to \mathbf{B}^m$ over the variable set $X = \{x_1, \cdots, x_n\}$, where $\mathbf{B}=\{0,1\}$. Here, $n$ is the number of variables (inputs), and $m$ is the number of functions (outputs).

### 2.1. Graph-based Representation of Boolean Functions

*Binary Decision Diagrams* (BDDs) have become the advanced structures in CAD of integrated circuits for representation and manipulation of Boolean functions.

BDD is a connected, directed acyclic graph, where:

**(i)** each non-terminal node corresponds to *Shannon expansion* $S$ of the function $f$ with respect to variable $x$ (incoming $edge$) into sub-functions (outgoing edges: $edge_l$ and $edge_r$): $f = \overline{x} \cdot f_{|x=0} \vee x \cdot f_{|x=1}$;

[*]Support from the University of Wollongong (AUSTRALIA) and Technical University of Szczecin (POLAND) is acknowledged

**(ii)** a starting node is called *root*; a terminal node is labeled with the leaf value and has no successors; a non-terminal node has exactly two successors.

A BDD is called *ordered* if the variable $x$ appears in the same order in each path from the root to a terminal node. The *size* of BDD (number of non-terminal nodes) may vary from linear to exponential depending on variables' order. A BDD is called *reduced* if it does not contain any nodes either with isomorthic sub-graphs or with both edges pointing to the same node. In our study, we always deal with reduced ordered BDDs (here, simply BDDs).

## 2.2. Information Measures in Digital Circuits

In order to quantify the content of information for a finite field of events $A = \{a_1, a_2, \cdots, a_n\}$ with probabilities distribution $\{p(a_i)\}$, $i = 1, 2, \cdots, n$, Shannon introduced the concept of entropy [4]. *Entropy* of the finite field $A$ is given by

$$H(A) = -\sum_{i=1}^{n} p(a_i) \cdot \log p(a_i), \quad (1)$$

where log denotes the base 2 logarithm function.

For two finite fields of events $A$ and $B$ with probability distribution $\{p(a_i)\}, i = 1, 2, \cdots, n$, and $\{p(b_j)\}$, $j = 1, 2, \cdots, m$, probability of the joint occurrence of $a_i$ and $b_j$ is joint probability $p(a_i, b_j)$, and there is conditional probability, $p(a_i|b_j) = p(a_i, b_j)/p(b_j)$. The *conditional entropy* of $A$ given $B$ is defined by

$$H(A|B) = -\sum_{i=1}^{n}\sum_{j=1}^{m} p(a_i, b_j) \cdot \log p(a_i|b_j). \quad (2)$$

In case of Boolean functions, we assume that the set of values of a function $f$ and the set of values of arbitrary variable $x$ are two finite fields [7]. We calculate the probability $p_{|f=b} = k_{|f=b}/k$, where $k_{|f=b}$ is the number of assignments of values to variables (patterns) for which $f = b$ and $k$ is the total number of assignments. Other probabilities are calculated in the same way.

**Example 1.** *For the function $f = \overline{x}_3 \cdot \overline{x}_2 \vee x_1$ with truth vector [10001111] according to Equations (1) and (2): $H(f) = -5/8 \cdot \log_2(5/8) - 3/8 \cdot \log_2(3/8) = 0.96\ bit$, $H(f|x_1) = -1/8 \cdot \log_2(1/4) - 3/8 \cdot \log_2(3/4) - 4/8 \cdot \log_2(4/4) - 0 = 0.41\ bit$. By the same computations we have $H(f|x_2) = 0.91\ bit$, $H(f|x_3) = 0.91\ bit$.*

## 3. BDD BASED TECHNIQUE FOR CALCULATION OF INFORMATION MEASURES

First and natural step of calculation of information measures is to determine the probabilities of a function and their subfunctions (see Equations (1) and (2)).

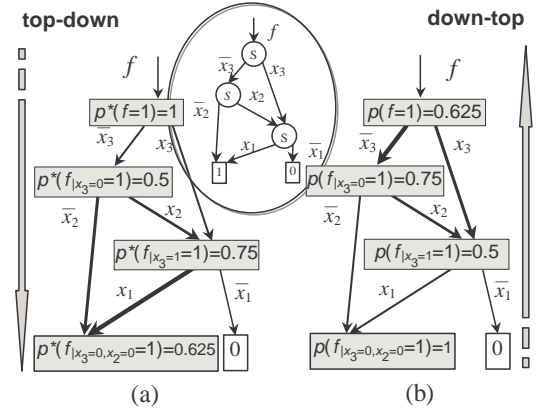

Figure 1: Calculation of output probabilities using BDD

### 3.1. Probability

First attempt to compute the output probabilities was proposed in [1], where an exact strategy based on BDDs is described which is linear in the size of the corresponding function graph (the size of the graph, however, may be exponential in the number of circuit inputs). The further extension of such BDD traversal in a down-top fashion was the probability assignment algorithm presented in [3]. In the case of completely specified Boolean function $f$: $p(x = 0) = p(x = 1) = 1/2$, and the output probability of the constant function "1" is 1 ($p(leaf|f=1) = 1$) and the output probability of "0" is 0 ($p(leaf|f=0) = 0$). Since each node of BDD is an instance of Shannon expansion, for a node representing a function $f$:

$$p(f) = 1/2 \cdot p(f_{|x=0}) + 1/2 \cdot p(f_{|x=1}). \quad (3)$$

An alternative is to use top-down strategy. The algorithm of probability assignment presented in [6] can be rewritten in our notations as: **(i)** $p(root) = 1$ for the nodes of $level = 0$; **(ii)** $p^*(node^{level+1}) = p^*(edge_1^{level}) + \ldots + p^*(edge_v^{level})$ for any node of the $level$ which has exactly $v$ incoming edges; **(iii)** $p(f = 1) = p^*(leaf|f = 1)$ and $p(f = 0) = p^*(leaf|f = 0)$.

**Example 2.** *Let us consider two strategies for calculation of output probabilities of the function $f$ from Example 1. We run top-down like BDD traversal with assigning $p^*(f = 1) = 1$ and obtain $p(f = 1) = p^*(leaf|f = 1) = 0.625$, as indicated in Figure 1(a). Down-top approach with assigning $p(leaf|f = 1) = 1$ gives us $p(f = 1) = p(root) = 0.625$, as shown in Figure 1(b).*

In our exploration, we utilize the modification of the first strategy where the actual output probability of the function $f$ is assigned to the root node. We extend the approach to calculating the output probabilities (Equation (3)) to the following recursive algorithm: $p(node) = $

$1/2 \cdot p(edge_l) + 1/2 \cdot p(edge_r)$, where $p(leaf|f=1)=1$ and $p(leaf|f=0)=0$. Thus, the output probability be $p(f=1)=p(root)$ and $p(f=0)=1-p(root)$. For generalization the coming next recursive algorithm is proposed:

$$p(node) = p(x=0) \cdot p(edge_l) + p(x=1) \cdot p(edge_r).$$

This algorithm allows to calculate conditional and joint probabilities which are needed for computation of conditional entropy according to Equation (2). Thus, for joint probability $p(f=1, x=1)$ it is needed to set $p(x=1)=1$ and $p(x=0)=0$ before BDD traversal. Such approach allowed us to develop a technique for calculation of the *whole range* of probabilities using only one BDD traversal.

**Example 3.** *The results of setting $p(x_2 = 0) = 1$ and $p(x_2 = 1) = 1$ for the function $f$ from Example 1 give us conditional probabilities $p(f=1|x_2=0) = 0.75$ and $p(f=1|x_2=1) = 0.5$.*

### 3.2. Shannon Information Measures

In [2] proposed an algorithm based on symbolic computations for exactly determining the entropies of digital circuits. It was reported that the technique is viable, as proved by the results obtained on a large set of mid-scale functions. This exact approach breaks down when applied to large examples.

Our technique for exact computation of Shannon information measures for Boolean functions represented in the form of BDDs exploits the following. Equation (1) is used for computing the entropy $H(f)$ of the function $f$ together with the described above algorithm for calculation of output probabilities. The conditional entropy $H(f|x)$ of the function $f$ with respect to the variable $x$ can be simplified using the theorem below:

**Theorem 1.** *The conditional entropy $H(f|x)$ can be calculated by the following equation:*

$$H(f|x) = p(x=0) \cdot H(f_{|x=0}) + p(x=1) \cdot H(f_{|x=1})$$

It means that for calculation of conditional entropy we need to compute the entropy of each sub-function. In this case probability must be assigned to every node in BDD in order to distribute the desired output probability to the root.

**Example 4.** *The entropy of the function $f$ from Example 1 be: $H(f) = 0.96$ bit. The conditional entropy of the function $f$ given $x_2$ be: $H(f|x_2) = 1/2 \cdot H(f_{|x_2=0}) + 1/2 \cdot H(f_{|x_2=1}) = 0.41 + 0.5 = 0.91$ bit (Figure 2). The same manipulation yields: $H(f|x_1) = 0.41$ bit and $H(f|x_3) = 0.91$ bit. The conditional entropy of the function $f$ given a set of variables $\{x_1, x_2\}$ be: $H(f|x_1 x_2) = 0.25$ bit.*

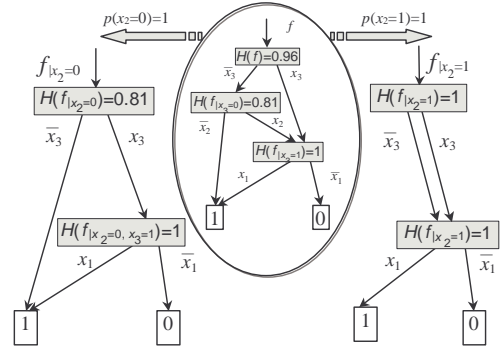

Figure 2: Computing information measures using BDD

```
Input BDD for the given function f
Output New order of the variables
Info^R(f)
{
  for(∀level : level = 1...n ) {
    for(∀x of the level ≥ level) {
      Calculate p(f = 1|x = 0) and p(f = 1|x = 1);
      Calculate H(f|x_i) using Theorem 1; }
    Reorder variables according to
     Equation (4):  H(f|x_{j_1}) ≤ ... ≤ H(f|x_{j_n});
    Rebuild BDD using x_{j_1} for current level; }
}
```

Figure 3: The sketch of the algorithm $Info^R$

### 4. BDD REORDERING ALGORITHM

We have realized the described above idea of calculation of information measures to obtain the whole range of information measures using only one BDD traversal, and we utilize these measures for BDD reordering. New BDD order is built taking into consideration the information criterion. The *criterion* to choose a decomposition variable $x$ for the arbitrary level of BDD is that the conditional entropy of the function $f$ with respect to this variable has to be minimal:

$$H(f|x) = min(H(f|x_i) \mid \forall \, x_i). \quad (4)$$

Figure 3 gives a sketch of the algorithm for BDD reordering.

### 4.1. Experiments

We have implemented the technique for calculation of information measures and BDD reordering algorithm within CUDD package [5]. We have run experiments on several sets of benchmarks (combinational and sequential circuits) selected from the ISCAS85 and ISCAS89 suite on Pentium III 650Mhz with 64Mb of memory.

In Table 1, we give the outline of the first set of experiments, where information measures have been calculated (entropy for the first output $f_0$ and conditional entropy with respect to the first $x_0$ and second $x_1$ inputs, all numbers in *bits*). In Table 2, we report the second set of investigations, where BDDs have been built for both combinational and sequential circuits using different reordering approaches (window, genetic, sifting and our $Info^R$). We observe that for some benchmarks algorithm $Info^R$ produces better results than other reordering methods.

| Test | Information measures | | |
|---|---|---|---|
| | $H(f_0)$ | $H(f_0\|x_0)$ | $H(f_0\|x_1)$ |
| C1908 | 1.00 | 0.017 | 1.00 |
| C3540 | 0.337 | 0.272 | 0.272 |
| C432 | 0.385 | 0.378 | 0.378 |
| C6288 | 0.811 | 0.50 | 0.50 |
| C880 | 0.544 | 0.406 | 0.406 |
| S1196 | 0.896 | 0.842 | 0.883 |
| S1512 | 0.954 | 0.905 | 0.749 |
| S344 | 0.811 | 0.50 | 0.749 |
| S4863 | 0.982 | 0.922 | 0.922 |
| S635 | 0.811 | 0.811 | 0.50 |

Table 1: Some values of information measures

| Test | Size [number of nodes] | | | |
|---|---|---|---|---|
| | Window | Genetic | Sifting | $Info^R$ |
| C17 | 11 | 11 | 11 | **9** |
| C1908 | **36007** | **36007** | **36007** | 65032 |
| C432 | **1733** | **1733** | **1733** | 2425 |
| C880 | 346660 | 346660 | **21965** | 319071 |
| S1423 | 98454 | 98454 | 98454 | **73680** |
| S298 | **125** | **125** | **125** | 175 |
| S349 | 206 | 206 | 206 | **141** |
| S382 | **168** | **168** | **168** | 171 |
| S820 | 2651 | 2651 | 2651 | **2283** |

Table 2: Different algorithms of BDDs reordering

### 4.2. Low-power Design: Case Study

For many consumer electronic applications, low average power dissipation is desirable, and for certain special applications low power dissipation is of critical importance.

**Example 5.** *According to the criterion (Equation (4)) and the values of conditional entropy (Example 4), the order of variables in BDD for the function $f$ will be $< x_1 x_2 x_3 >$. The power dissipated by circuits derived from BDDs[1]:*

---
[1] We use *SIS Release 1.2., UC Berkeley, 1994*, as an environment for modelling of power dissipation (function **power_estimate**).

| Variables' order | | $P, \mu W$ |
|---|---|---|
| $< x_1 x_2 x_3 >$ or $< x_1 x_3 x_2 >$ | | 31.87 |
| $< x_2 x_1 x_3 >$ or $< x_3 x_1 x_2 >$ | | 45.0 |
| $< x_3 x_2 x_1 >$ or $< x_2 x_3 x_1 >$ | | 37.50 |

*The power consumption for best case is 1.5 times smaller than for the worst case taking into account the same size of BDDs for both cases.*

## 5. CONCLUDING REMARKS

The purpose of this paper was to investigate possibility to calculate different information measures using BDD. We have presented the BDD based technique for computing of information measures from a theoretical as well as practical perspective. We can indicate what this problem can be easily solved when Boolean function is represented as BDD.

We have proposed reordering algorithm based on information criterion which provides an improvement in BDD size for some benchmarks in contrast to existing reordering methods. The BDD reordering using information measures was shown to be interesting for further applying in different problems, especially in the area of low-power design of digital circuits derived from BDD. It is hoped that this paper will help to widen the understanding of practical applications of information theoretical concepts.